\documentclass[twocolumn,superscriptaddress,prl,floatfix]{revtex4}
\usepackage[dvipdfm]{graphicx}
\usepackage{color}
\usepackage{soul}
\usepackage{bm}

\begin{document}

\title{Emergence of ferroelectricity and spin-valley properties
in two-dimensional  honeycomb binary compounds.}

\author{Domenico Di Sante} \affiliation{Consiglio Nazionale delle Ricerche
(CNR-SPIN), Via Vetoio, L'Aquila, Italy} \affiliation{Department of Physical and
Chemical Sciences, University of L'Aquila, Via Vetoio 10, I-67010 L'Aquila,
Italy}\email{domenico.disante@aquila.infn.it}

\author{Alessandro Stroppa} \affiliation{Consiglio Nazionale delle Ricerche
(CNR-SPIN), Via Vetoio, L'Aquila, Italy}\email{alessandro.stroppa@aquila.infn.it}

\author{Paolo Barone}
\affiliation{Consiglio Nazionale delle Ricerche
(CNR-SPIN), Via Vetoio, L'Aquila, Italy}

\author{Myung-Hwan Whangbo} \affiliation{Department of Chemistry, North Carolina
State University, Raleigh, North Carolina 27695-8204, U.S.A.}

\author{Silvia Picozzi} \affiliation{Consiglio Nazionale delle Ricerche
(CNR-SPIN), Via Vetoio, L'Aquila, Italy}

\date{\today}

\begin{abstract}

By means of density functional theory calculations, 
we predict that
several two dimensional AB binary monolayers, where 
A and B atoms belong to group IV 
or III-V, are ferroelectric. Dipoles arise from the 
buckled structure, 
where the A and B ions are located on the sites of a 
bipartite corrugated 
honeycomb lattice with trigonal symmetry. We discuss the emerging valley-dependent 
properties and the coupling of spin and valley physics, which arise from the loss of
inversion symmetry, and explore the interplay between
 ferroelectricity and Rashba spin-spitting phenomena. 
We show that valley-related 
properties 
originate mainly from the binary nature of AB monolayers, while the Rashba 
spin-texture developing around valleys is fully controllable and switchable 
by reversing the ferroelectric polarization.

\end{abstract}


\maketitle

\section{INTRODUCTION}
A wide range of modern electronic applications are based on
the charge and spin degrees of freedom (DOF) of electrons. Two-dimensional (2D)
atomic crystals with honeycomb lattice, such as graphene and molybdenum disulfide
(MoS$_2$) monolayer, have recently been the object of intense 
research activities due to the additional valley DOF of carriers that might be useful in next-generation
electronics applications
\cite{graphValley,graph1,graph2,graph3,mos2_1,mos2_2,mos2_3,mos2_4,mos2_5,mos2_6,mos2_7,mos2_8,mos2_spinvalley}. 
In the 2D semimetal graphene \cite{graphValley,graph1,graph2,graph3}, the $\pi$ and $\pi^{*}$ bands
linearly cross at K and -K points of the hexagonal Brillouin zone (BZ),
 implying that charge carriers behave like
massless Dirac fermions;
 at the same time, when spin-orbit coupling (SOC) is taken into account, 
the electrons experience opposite effective magnetic fields with equal 
magnitude at the K and -K valleys (related by time-reversal symmetry).
In principle, valley DOFs in graphene could be exploited for 
valley-dependent electronics
and optoelectronics, but their control by electrical and optical
means is difficult due to the inversion symmetry of 
the graphene crystal structure, preventing the appearance of 
valley-contrasting properties\cite{Xiao_2012,mos2_8}. 
On the other hand, the 2D MoS$_2$ semiconductor 
\cite{mos2_1,mos2_2,mos2_3,mos2_4,mos2_5,mos2_6,mos2_7,mos2_spinvalley,mos2_8} 
monolayer has no inversion
symmetry and displays a direct band gap at the K and -K valleys, 
enabling optical pumping of valley-polarized carriers by shining the monolayer with circularly polarized light\cite{mos2_1,mos2_2}. In addition, when
electrically biased, electrons from the K and -K valleys of MoS$_2$ monolayer
experience opposite Lorentz-like forces giving rise to a valley Hall effect
(VHE). To measure the VHE, it is necessary to irradiate the layer with
circularly polarized photons (with the electric field applied parallel to the
layer) so that electrons are excited only from valley K (or -K) 
thus breaking time-reversal symmetry\cite{mos2_7,mos2_8}.

\begin{table*}[!t]
\centering
\caption{Buckling angle $\theta$ (deg), buckling height $h$ (\AA), band gap
E$_g$ (eV), barrier height E$_a$ (eV/fu), polarization P ($10^{-12}~C/m$) and valence $\Delta E_{VB}$
and conduction $\Delta E_{CB}$ spin-splittings (meV) at the K point of
buckled AB monolayers.}
\label{tab1}
\begin{tabular}{p{1cm}ccccccc}
\hline\hline
     & $\theta$ (deg) & $h$ (\AA)  & E$_g$ (eV) & E$_a$ (eV/fu) & P ($10^{-12}C/m$) & $\Delta E_{VB}$ (meV) & $\Delta E_{CB}$ (meV)  \\
\hline
SiGe & 105.1 & 0.60 &  0.02     & 0.16  & 0.88 &  25.3  &   5.8  \\
SiSn & 105.4 & 0.72 &  0.98     & 0.23  & 4.22 &  79.8  &  38.0  \\
GeSn & 108.2 & 0.81 &  0.21     & 0.39  & 3.22 &  70.0  &  19.1  \\
AlSb & 105.2 & 0.68 &  1.43     & 0.14  & 7.82 &  20.0  &  50.8  \\
GaP  & 101.3 & 0.44 &  2.16     & 0.05  & 9.24 &   7.4  &   5.3  \\
GaAs & 105.3 & 0.63 &  1.74     & 0.18  & 9.07 &  14.9  &  33.8  \\
InP  & 102.4 & 0.53 &  1.33     & 0.07  & 11.45 &  21.0  &  10.4  \\
InAs & 105.9 & 0.70 &  0.87     & 0.22  & 11.10 &  36.4  &  45.8  \\
InSb & 107.5 & 0.83 &  0.69     & 0.30  & 8.30 &  57.2  & 100.9  \\
\hline
\hline
\end{tabular}
\end{table*}

\begin{figure}[!b]
\centering
\includegraphics[width=\columnwidth,angle=0,clip=true]{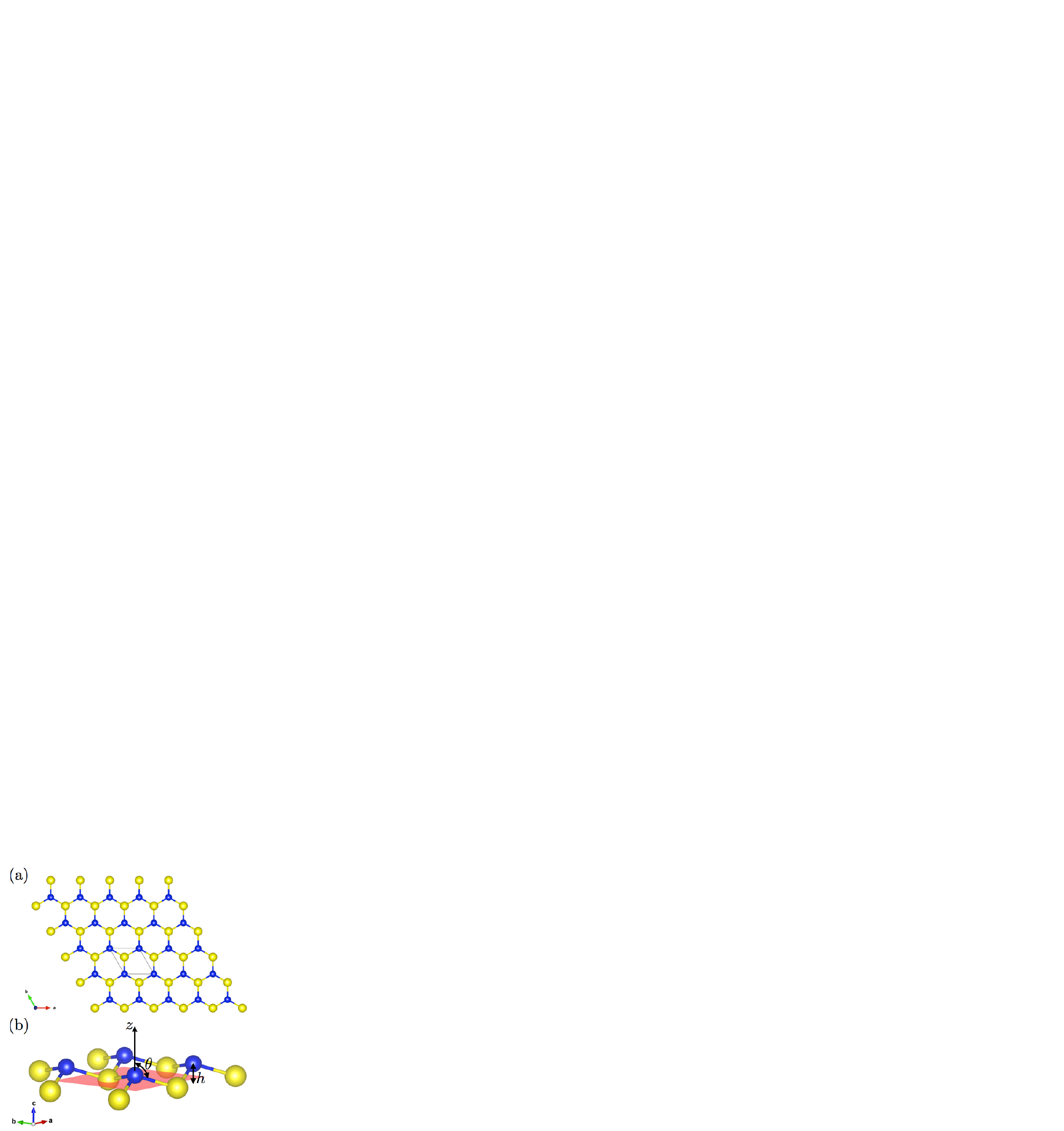}
\caption{(AB = SiGe, SiSn, GeSn, AlSb,
GaP, GaAs, InP, InAs, InSb) (a) Top view of a buckled AB monolayer along the
c-direction. (b) 3D view of a buckled AB monolayer,
where the trigonal sheet of A is separated by a distance $h$ from that of B.
The buckling angle $\theta$ is defined as the angle between an A-B bond and the $z$ direction
normal to the plane. AB$\rightarrow$BA conversion and the consequent ferroelectric
switching occurs when $\theta$ crosses $\pi/2$.}
\label{fig1}
\end{figure}

Using first-principle calculations,
Ciraci \textit{et al.} investigated
two-dimensional honeycomb structures of group-IV elements and their 
binary compounds as well as the compounds of group III-V elements\cite{buckledAB}. 
It was found that  
buckled AB monolayers with trigonal
symmetry (\textit{e.g.}, group IV binary monolayers SiGe, SiSn, 
GeSn and group
III-V binary monolayers AlSb, GaP, GaAs, InP, 
InAs, InSb) \cite{buckledAB}, in which a
trigonal sublattice of A ions is separated from that of B ions (Fig. \ref{fig1}a), 
are more stable with respect to a planar geometry. 
The tendency to a buckled geometry was explained in terms 
of the destabilization of the $\pi$ bonds in $sp^{2}$ hybridization 
due to the increase of the bond length between the two atoms A and B,
as it happens in silicene and 
germanene compared to graphene.\cite{silicene}
As a result of the buckling, the $sp^{2}$ states 
dehybridize and the $s$,$p_{x}$,$p_{y}$ orbitals combine with 
$p_{z}$ orbitals to form new $sp^{3}$-like orbitals. Three of these 
$sp^{3}$ states form covalent bonds with three nearest-neighbor atoms, 
while one $sp^{3}$-like orbital directed upward perpendicular
to the atomic plane forms a weak bond with the adjacent $sp^{3}$-like 
orbital directed downward.
Interestingly, silicene and germanene are expected to display
valley-contrasting properties analogous to graphene 
as soon as the inversion symmetry is broken, e.g., by applying an external electric field.\cite{silicene}
Therefore,  other buckled honeycomb lattices
are expected to display similar
valley-dependent properties.  Furthermore, 
in  \textit{buckled} trigonal structure, 
one can introduce a {\it sublattice pseudospin} $\zeta$ describing 
the binary layer DOF
\cite{mos2_8}: 
the pseudospin $\zeta$ up (down) refers 
  to the state where the charge carrier is located 
 in the upper (lower) layer, or equivalently in the A (B) sublattice. 
Therefore, a pseudospin polarization would directly 
correspond to an electrical polarization.  In fact,
buckled AB monolayers have no
inversion symmetry and actually belong to the \textit{polar} space group $P3m_1$, with the polar axis
perpendicular to the layer (Fig. \ref{fig1}b), possibly leading to a ferroelectric (FE) state; 
in addition to the emergence of valley-contrasting physics,
 therefore, they can also display a Rashba effect\cite{Rashba}, 
which would likely be coupled to and controllable with the
 FE polarization. If this is the case, 
it would be possible to act on
the spin DOFs of these valley-active systems by  reversing the FE
polarization of the monolayer \cite{GeTe}. 
We explore this possibility in the present work, where we theoretically predict, by density functional calculations 
complemented with  model Hamiltonian analysis, 
that the  buckled group IV and group III-V 
binary monolayers with trigonal symmetry\cite{buckledAB} are 
2D ferroelectrics with spin-valley coupling and Rashba effects in their
electronic structure. 
We show that SOC and a bipartite honeycomb lattice
with different A and B atoms are at the origin of a spin-valley-sublattice 
coupling which is responsible for a valley-dependent Zeeman-like splitting
at the K and -K valleys, while a nonvanishing buckling is responsible for an
in-plane Rashba spin-texture that is controllable by electric field.   

From the experimental point of view, several top-down and bottom-up methods have been proposed and devised for synthesizing 2D materials, as reported in recent review papers\cite{rev_exp1, rev_exp2}. Specifically, both silicene and germanene
have been proven to exist as a monolayer when grown on selected metal substrates. Although
bulk Si cannot form a layered phase like graphite, experiments 
of surface-assisted epitaxial growth show the presence of nanoribbons 
of silicene on Ag(110)\cite{silicene_exp1} and 2D monolayers with buckled honeycomb structure on
Ag(111)\cite{silicene_exp, silicene_exp2} and Ir(111)\cite{silicene_exp3}. Similarly, successful attempts to grow 2D germanium sheets with a honeycomb structure on a platinum(111) and gold(111) template have been reported very recently\cite{germanene_exp1,germanene_exp2}. 
As for binary compounds, to the best of our knowledge, no other 2D monolayers beside boron nitride
have been synthesized yet\cite{atlas}, even though almost 2D nanoflakes of SiC with thickness of the order of 1 nm have been obtained by means of solution-based exfoliation of SiC crystals\cite{SiC_exp}.
We hope that our work can stimulate further experimental 
work to fill this gap.

\section{METHODS}
Our density functional calculations for buckled AB monolayers were carried out
using the projector augmented wave (PAW) method implemented in VASP using the
PBEsol functional \cite{paw,vasp,pbe}. $d$-electrons are included in the  valence in the PAW potentials. Test calculations were also performed using the
Heyd-Scuseria-Ernzerhof (HSE) screened hybrid density functional \cite{hse}. The atom
positions were optimized until the residual forces were smaller than 0.001 eV/\AA\,
with the plane-wave cutoff energy of 500 eV and a set of 8$\times$8$\times$1 k-points for the
irreducible BZ. Repeated image interactions were made negligible by including a
vacuum layer of 26 \AA\, in the simulations. The FE polarization is calculated
using the Berry's phase method \cite{berry}, and SOC is self consistently taken into account.

The first-principle Berry curvature
is calculated according to the usual linear response Kubo-like formula
\begin{eqnarray}
\label{berry_formula}
\Omega({\bf k})&=&\sum_n f_n\Omega_n({\bf k}) \nonumber \\
\Omega_n({\bf k})&=&-2 Im \sum_{m\ne n}\frac{\langle u_{n{\bf k}}|v_x|u_{m{\bf k}}\rangle \langle u_{m{\bf k}}|v_y|u_{n{\bf k}}\rangle}{(E_{m{\bf k}} - E_{n{\bf k}})^2},
\end{eqnarray}
where $f_n$ is the Fermi distribution function, $v_{x,y}$ is the velocity
operator, and $u_{n{\bf k}}$ is the lattice-periodic eigenvector with eigenvalue
$E_{n{\bf k}}$ of the Fourier transformed Wannier Hamiltonian as calculated by
projecting the DFT Hamiltonian onto a Wannier basis \cite{wannier90}.

\section{RESULTS}

\begin{figure}[!h]
\centering
\includegraphics[width=\columnwidth,angle=0,clip=true]{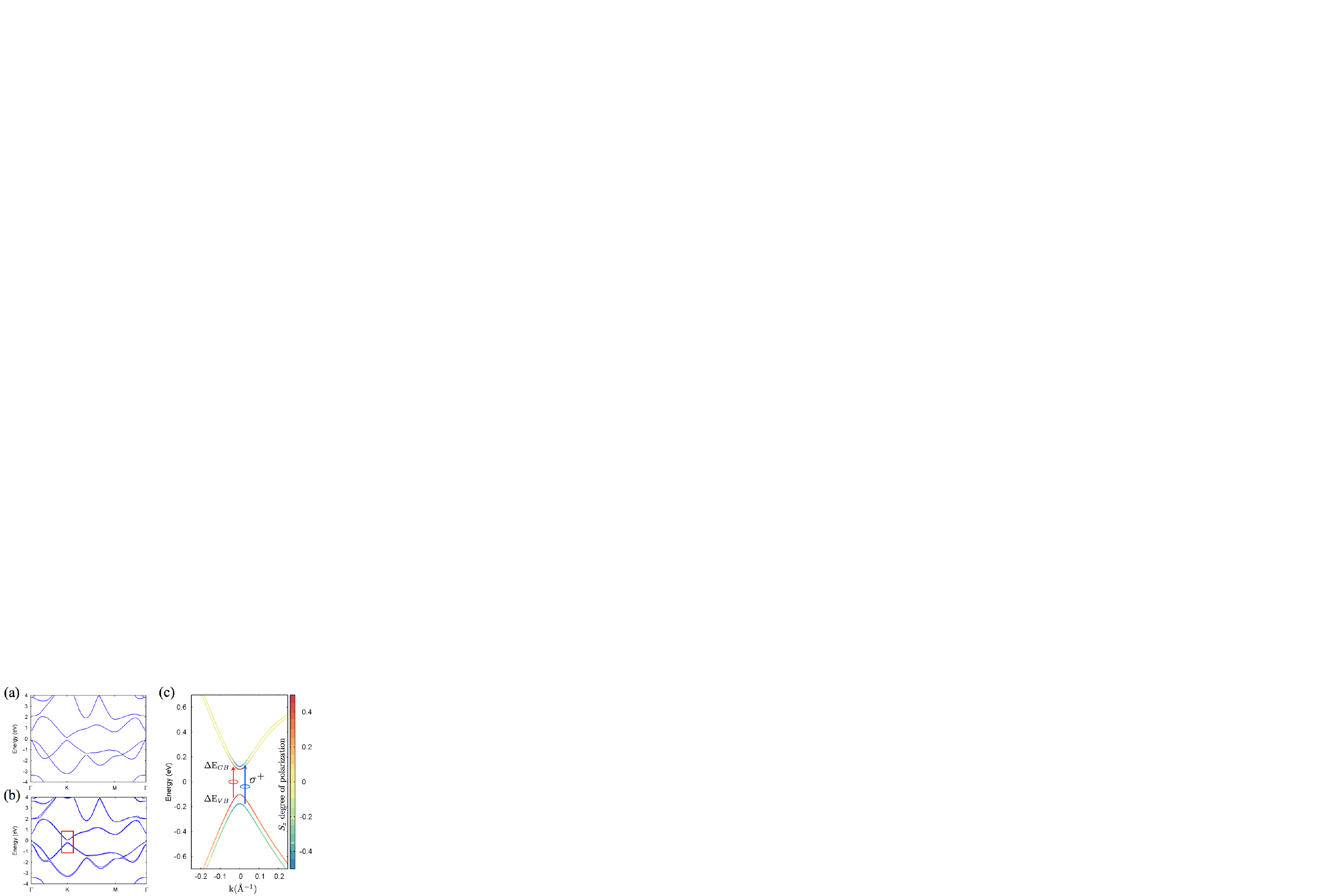}
\includegraphics[width=\columnwidth,angle=0,clip=true]{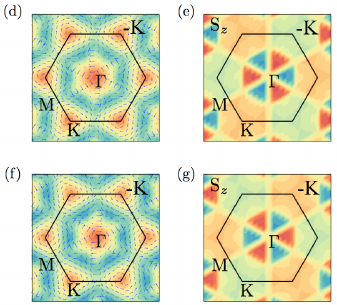}
\caption{(a) Band structures along the 2D hexagonal Brillouin zone (BZ) (a) without and
(b) with SOC. In (b) the red panel is an eyeguide for the zoomed-in view in (c), where low
energy bands around the Fermi level in the vicinity of K are reported. Color
scale refers to the out-of-plane $S_z$ spin polarization, while the red and blue
arrows, referring to positive and negative $S_z$ spin-component gaps respectively, denote the
valley and spin optical transition selection rules for circularly polarized
light needed at K for optical interband excitations. (d,e) In-plane and
out-of-plane spin textures of the upper and (f,g) lower spin-split VBs for the
whole BZ. In (d,f) the color coding refers to the wavevector-dependent energy highlighting the
$C_{3v}$ symmetry, while the color scale for the $S_z$ degree of polarization in
(e,g) is consistent with (c).}
\label{fig2}
\end{figure}

Starting from the buckled compounds listed in ref.\citenum{buckledAB}
we have calculated 
the buckling heights h (\AA), band gaps E$_g$ (eV), barrier heights E$_a$ (eV
per formula unit) estimated 
as the energy difference between the FE buckled structure and the paraelectric
planar one, FE polarizations P ($10^{-12}~C/m$) and spin-splittings $\Delta E_{VB}$ ($\Delta E_{CB}$) (in meV)
at the K point for valence (conduction) bands, see Table \ref{tab1}. 
In this work, we follow previous theoretical studies\cite{buckledAB} and assume that atoms A and B belong to
the ordered bipartite lattice shown in  Fig. \ref{fig1}a),
\textit{i.e.}, we neglect the role of disorder\cite{disorder} ;
indeed this effect would require a separate analysis, beyond the 
purpose of the present study.
First we note that if A, B atoms belong to the same group 
(\textit{i.e.}, IV group), 
the A-B bond is polar due to the electronegativity difference 
between A and B, since the electronegativity within a given family of elements decreases on going from the top to a lower period. On the other hand, when A, B atoms belongs to III and V group, respectively, 
the electronegativity difference becomes even more pronounced (the electronegativity 
increasing from left to right along the period). The trigonal symmetry and the buckled honeycomb structure imply a local uniaxial dipole 
moment along the 3-fold rotation axis. By analogy with the Ising model for uniaxial ferromagnets, which is well known to display a phase transition even in the two-dimensional lattice\cite{onsager_ising}, a FE phase transition is indeed possible, where the two symmetry-equivalent energy minima with opposite polarity are obtained by reversing the buckling angle. 
The estimated FE polarizations of the group IV
binary AB monolayers are significantly large, 
with calculated typical values of
the order of 1-4 $\times 10^{-12}~C/m$ while the group
III-V binary AB monolayers have a larger FE polarization, 
because of the larger electronegativity difference and larger dipole moment carried by each A-B
bond. Typical values in this case are around 10
$\times 10^{-12}~C/m$. It is interesting to note that 
the calculated values for the estimated FE polarizations are always one order of magnitude larger than those measured in 2D freely-suspended FE smectic-C films\cite{meyer_1979} and nematic monolayers\cite{tabe_2003}, showing $P\sim 10^{-13}~C/m$.
Evaluation of Born effective charges $Z^*$ 
confirms the estimated values of FE
polarization, at the same time providing an estimate of the depolarization
field, which is expected to significantly affect the FE properties of
thin films. In fact, the electrostatic energy of the depolarization field is
proportional to the the square of $Z^*$  and inversely proportional to the
electronic polarizability of the material; since the latter is 
almost constant for all the considered systems, {\it i.e.} $\varepsilon^{\infty}_{zz}\sim 1$,
while $Z^*$ is significantly 
small (for the $zz$ component of the charge tensor we calculate $Z^*_{zz}$ $\sim 0.05e$ for group IV and $\sim 0.1e$
for group III-V binary AB monolayers), the depolarization field can be expected to be weaker
than that preventing ferroelectricity in ultrathin films of ferroelectric
perovskites, supporting the feasibility of stable FE distortions.
Eventually,
the energy barriers E$_a$ calculated for the AB$\rightarrow$BA
conversion are comparable with those estimated for conventional FE oxides, such as PbTiO$_3$ for which E$_a\sim 0.1-0.2~eV/$formula unit, as well as with that of the recently predicted 1$T$ FE phase of MoS$_2$ \cite{mos2_ferro},
suggesting that the polarization reversal could be experimentally accessible. Inelastic electron excitations from a STM tip could be also used to switch between the two FE phases, as recently proposed for bistable molecular switches\cite{fu_pccp_2010,repp_prl_2012}.

We discuss then the electronic band structures of a representative example, GeSn, calculated
without and with SOC and shown in Fig. \ref{fig2} (a-b). Band structures for all the other compounds are displayed 
in the Supplementary Material.
The band gap opens
at the K point, where both the valence (VB) and conduction (CB) bands are split by
SOC, as highlighted in Fig. \ref{fig2}(c). It is also clear from
Fig. \ref{fig2}(c) that electrons around the K   valley  feel a strong Zeeman-like
magnetic field, which is responsible for a valley-dependent 
out-of-plane spin polarization both in
the VB and in the CB \cite{mos2_5}. The same holds for the -K valley. 
Due to the time-reversal symmetry, the system remains overall non-magnetic, 
with opposite out-of-plane spin polarization at time-reversed K and -K points, 
as clearly shown in Fig. \ref{fig2} (e),(g).
Therefore, the considered buckled FE monolayers show coupled spin 
and valley 
physics, thus possibly allowing the  spin and valley control similar  to
 layered transition metal dichalcogenides.
As for MoS$_2$ and other group-VI dichalcogenides, 
the valley Hall effect should be accompanied by a spin Hall effect for 
both the electron and hole-doped systems, whose robustness can be deduced by the expected long relaxation time of spin and valley indices\cite{mos2_spinvalley}. Because of the spin-valley coupling and the valley-contrasting spin splittings $\Delta E_{VB}$ and $\Delta_{CB}$, the energy conservation would imply a simultaneous flipping of spin and valley indices; since two adjacent valleys are separated by a wave vector comparable with the size of the Brillouin zone, such simultaneous flipping would require atomic scale (magnetic) scatterers. In the absence of such scatterers, both holes and electrons are expected to display long spin and valley lifetimes, therefore
allowing  robust Hall effects around both the valence and conduction-band edges \cite{mos2_spinvalley}.
The spin textures of the two spin-split VBs calculated for the
whole BZ are presented in 
Fig. \ref{fig2} (d-g).  
The typical Rashba spin patterns are 
clearly observed around K and -K valleys, 
with the in-plane spin components rotating clockwise or 
counterclockwise in spin-split bands, as shown in Fig. \ref{fig2}
 (d),(f) for upper and lower VBs.
Interestingly, while the out-of-plane spin components show opposite polarizations at time-reversed valleys, the in-plane spin components display the same chirality at K and -K points in a single VB; both the Rashba-like chirality and the valley-dependent magnetic moments appear then to be reversed in the other spin-split VB.
A similar behaviour is found in the spin-split CBs, while the VB maximum (VBM) and CB minimum (CBM) show same chiralities and out-of-plane polarization direction. 
When the FE polarization is switched by reversing the 
buckling, the in-plane spin-texture chiralities are fully reversed, suggesting the possibilities to control the Rashba effect by exploiting the FE properties of binary monolayers. On the other hand, the
 out-of-plane spin polarization remains exactly 
the same in opposite FE states. Interestingly, 
the valley-dependent spin polarization survives 
even when the buckling is completely suppressed 
as in the flat graphene-like structure, as a consequence 
of the non-centrosymmetric, albeit nonpolar, character 
of the planar honeycomb structure with binary composition. 

\begin{figure*}[!ht]
\centering
\includegraphics[width=\textwidth,angle=0,clip=true]{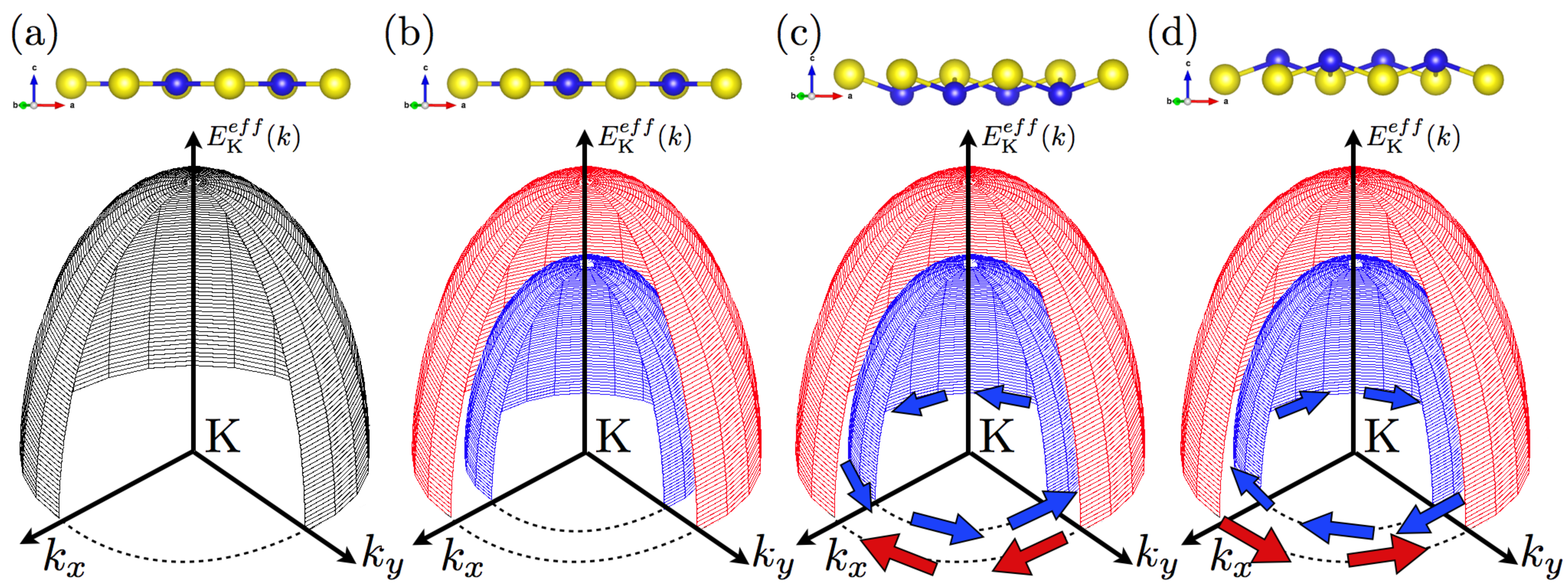}
\caption{(a-b) Valence bands (VBs) at the valley without (a) and with (b) SOC in planar AB monolayers. 
Red and blue colored bands in (b - d) refer to the presence of positive and negative out-of-plane spin polarizations,
respectively. (c-d) VBs and relative in-plane spin-texture for positive and negative
buckling. Red and blue colored bands still refer to out-of-plane spin polarization, while the arrows
are the spin projection on the $k_x,k_y$ plane.}
\label{fig3}
\end{figure*}

To understand the origin of this exotic spin and valley physics, 
and its interplay with 
the intrinsic FE polarization in buckled trigonal monolayers, we 
now investigate the microscopic mechanisms underlying the low-energy properties 
around the Fermi level at the K and -K valleys. In the absence of buckling ($\theta=\pi/2$)
and SOC, the low-energy effective Hamiltonian around the K point describing
 low-energy states with mainly $p_z$ character 
reads
\begin{eqnarray}
\label{form1}
H_{\text K}^{(0)} = -m\zeta_z +v_F(\tau k_x \zeta_x-k_y\zeta_y) 
\end{eqnarray}
where ${\boldsymbol \zeta}$ is the sublattice pseudospin trasforming like Pauli matrices, $\tau=\pm 1$ is a valley index for $\pm K$ points, $v_F$ is the Fermi velocity, and the first "mass" term originates from the different ions located in
A and B sublattices. Contrary to the case of graphene,
silicene and other group-IV binary monolayers \cite{graph2,graph_tb,silicene_tb,silicene_electricfield}, 
the additional mass term leads to an intrinsic gap in the
energy spectrum, which opens at the Dirac point
(the effective parameters are estimated in a tight-binding framework and given in the Supplementary Material). 
The VBs and CBs remain spin-degenerate, as sketched in 
Fig. \ref{fig3}(a) for VBs. When SOC is turned on in the planar structure ($\theta=\pi/2$), additional terms appear in the effective Hamiltonian (\ref{form1}),
namely
\begin{eqnarray}
\label{form2}
H_{\text K}^{(1)} = H_{\text K}^{(0)} -\tau\sigma_z\lambda_{so}^+\zeta_z -\tau\sigma_z\lambda_{so}^-
\end{eqnarray}
where $\lambda_{so}^\pm=(\lambda_{so}^A\pm\lambda_{so}^B)/2$ 
are effective material-dependent parameters arising from the interplay of atomic SOC constants $\xi_{A/B}$,
local orbital energies $\Delta_{A/B}$ and hopping integrals $V_{sp\sigma}$,$V_{pp\sigma}$ and $V_{pp\pi}$, which in turn
depend on geometrical factors such as the buckling angle $\theta$ (see Supplementary Material). The presence of SOC, therefore, introduces
an effective Zeeman-like valley-dependent magnetic field $B^{eff}=-\tau\lambda_{so}^-$, which removes 
the spin-degeneracy without mixing spin up and spin down states, leading to spin-split VBs and CBs with energies $E_\sigma=-\sigma\,\tau\lambda_{so}^-\,\pm\,\sqrt{(m+\sigma\tau\lambda_{so}^+)^2+v_F^2 k^2}$ and a net out-of-plane
spin polarizations at the K valleys, as sketched in Fig. \ref{fig3}(b). The VB and CB spin-splittings are listed in Fig. \ref{tab1} for all considered binary monolayers, and shown in 
Fig. \ref{fig2}(c) for the representative case of GeSn.
Additionally, the mass term acquires a spin-valley-sublattice contribution $\tau\sigma_z\zeta_z\lambda_{so}^+$ which indeed guarantees the coupling between the spin and valley physics.
Interestingly, the additional coupling terms experienced by a given sublattice
originate from the atomic SOC of the other sublattice mediated by the hopping
interactions --- in fact, carriers in the A sublattice feel the atomic SOC of B
ions through  the term $\lambda^B_{so}\sigma_z$, and vice versa. This is
reflected in the different size of spin-splitting gaps in VBs and CBS.
As shown in Fig. \ref{fig2}(c) for the representative case of GeSn,  
 the spin-splitting is larger at VBs rather than at CBs, despite the fact that VBs and CBs show predominant Ge and Sn characters, respectively, as a consequence of the larger electronegativity of Ge. Na\"ively, one 
would expect a larger spin-splitting at CBs than at VBs, since the atomic SOC constant of Sn is larger than that
of Ge, $\xi_{Sn}>\xi_{Ge}$.\cite{atomicSOC}  Indeed, the opposite is observed, since carriers with
a predominant Ge character experience the SOC-induced interaction coming from the Sn ions, and vice versa. 
The same holds for all other compounds, as can be inferred looking at values reported in Table 1. 
It is important to stress that such spin-splitting effects arise uniquely from
the binary composition of the monolayers, implying $\lambda^-_{so}\neq 0$. 
The spin-valley coupling, emerging already in the planar noncentrosymmetric binary monolayer, is therefore independent of the buckling distortion; in fact, the $\lambda_{so}^{A/B}$ are even functions of the buckling angle $\theta$, implying that valley-constrating properties such as the out-of-plane spin polarization are not expected to qualitatively change under FE distortions.
Furthermore, since the valley-dependent coupling terms do not mix the spin-up
and spin-down components, $\sigma_z$ remains a good quantum number, analogously
to what happens in MoS$_2$ monolayers. 

One can easily evaluate the Berry
curvature of the Bloch electrons, associated with the valley and spin Hall effect:
\begin{eqnarray}\label{berry}
\Omega_\sigma=\mp\tau\,\frac{2v_F^2(m+\sigma\tau\lambda_{so}^+)}{\left[v_F^2 k^2+(m+\sigma\tau\lambda_{so}^+)^2\right]^{3/2}},
\end{eqnarray}
where the $\mp$ sign refers to CB and VB, respectively, as well as the coupling strength with optical fields of $\hat{\sigma}^{\pm}$ circular polarization:
\begin{eqnarray}\label{osc_str}
\vert\mathcal{P}_{\pm}(k)\vert^2\propto\left(1\pm\tau\frac{m+\sigma\tau\lambda_{so}^+}{\sqrt{v_F^2 k^2+(m+\sigma\tau\lambda_{so}^+)^2}} \right),
\end{eqnarray}
which display the same form found for MoS$_2$ monolayers, leading to similar expectations
about the robustness of (valley and spin) Hall physics  and
optoelectronic effects\cite{mos2_spinvalley}.
In particular, the Berry curvature 
shows opposite sign in different valleys, while the interband optical
transitions are found to be uniquely coupled with $\hat{\sigma}^{+}$
($\hat{\sigma}^{-}$) circularly polarized optical field at the K (-K) valley,
the valley optical selection rules being also spin-dependent as shown in
Fig. \ref{fig2} (c). It is also clear from Eqs. (\ref{berry}), (\ref{osc_str}) that
the strength of such spin-valley physics is mainly governed by the
spin-valley-sublattice coupling parametrized by $\lambda_{so}^+$, even though
the valley-contrasting spin splitting of both VBs and CBs is due to the
effective magnetic field $\tau\lambda_{so}^-$.

\begin{figure}[!b]
\centering
\includegraphics[width=\columnwidth,angle=0,clip=true]{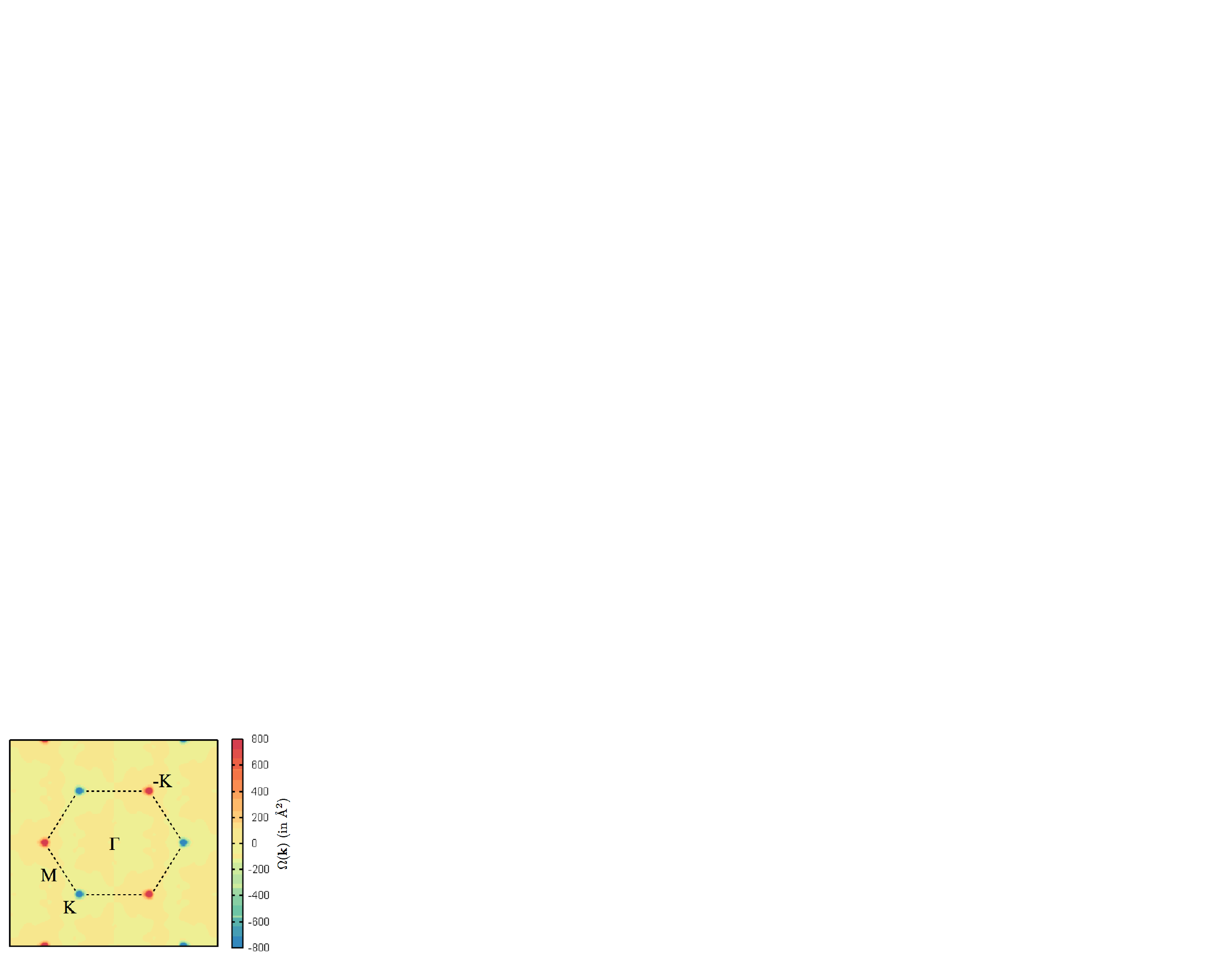}
\caption{Momentum resolved first-principles Berry curvature (in \AA$^2$) 
calculated for GeSn monolayer as a representative example.}
\label{fig_berry}
\end{figure}

When ferroelectricity sets in, lowering the trigonal symmetry from $C_{3h}$ to $C_{3v}$ and leading to a finite
buckling $\theta\ne\pi/2$, an intrinsic Rashba term must be added in the effective Hamiltonian Eq. (\ref{form2}), namely:
\begin{eqnarray}
\label{form4}
H_{\text K}^{(2)} = H_{\text K}^{(1)}-(\lambda_R^+ \zeta_z+\lambda_R^-)(k_y\sigma_x - k_x\sigma_y)
\end{eqnarray}
where $\lambda^{\pm}_R=(\lambda^{A}_R\,\pm\,\lambda^{B}_R)/2$ and $\lambda^{A/B}_R$ is a
complex material-dependent parameter which is odd under the switching of the buckling angle
 (see Supplementary Material). This Rashba coupling term
 gives rise to in-plane circularly rotating spin-texture around each K
valley, with opposite chiralities at spin-split VBs (CBs) as shown in Fig. \ref{fig3}(c)), while substantially not affecting the out-of-plane spin polarization. Neglecting $\lambda^{-}_R$ for the sake of simplicity, the expectation value for in-plane spin polarizations reads $\langle \sigma_{x} \rangle\propto -\lambda_R^+ k_y$ and $\langle \sigma_{y} \rangle\propto \lambda_R^+ k_x$ for the VBM and CBM (the overall chirality being fully reversed in the VBM-1 and CBM+1 branches), i.e., the typical Rashba-like behaviour, which appears to be valley-independent, in agreement with our DFT calculations.
Since $\lambda^{A/B}_R$ is an odd function of the buckling angle with respect to planar structure $\theta=\pi/2$, 
the buckling reversal, i.e., the switching of the ferroelectric
polarization, leads to a complete reversal of the in-plane Rashba spin-texture chirality only, while the VBs and 
CBs preserve their out-of-plane spin-polarization, as schematically shown in Fig. \ref{fig3}(c-d).
This behaviour perfectly agrees with our first-principles calculations, since its origin lies
in the opposite $\theta$-dependence of spin-valley ($\lambda_{so}^{A/B}$) and intrinsic Rashba
($\lambda^{A/B}_R$) coupling constants at, and around, the K point (note that $\lambda^{A/B}_R\neq 0$ only
when $\theta\ne\pi/2$ and the k-vector differs from K \cite{silicene_tb}). It is
worthwhile to notice that, although the Rashba-like coupling term causes a
mixing of spin-up and spin-down states, the valley physics appears to be robust
to buckling distortion. In fact, the Berry curvature has been evaluated from
first-principles for the representative GeSn with buckled honeycomb structure,
and it has been found to display opposite sign at K and -K valleys, as shown in
Fig. \ref{fig_berry}.

\section{CONCLUSIONS}
Our first-principles calculations predict a
spontaneous FE polarization in buckled
group IV and group III-V binary monolayers with trigonal symmetry.
Unlike the case of graphene, silicene, germanene and other single atomic type
monolayers, the presence of a diatomic basis that breaks the inversion symmetry  --- even in the planar geometry --- leads to the emergence of Zeeman-like spin-split bands with
coupled spin-valley physics analogous to the MoS$_2$ case. At the same time, it is
mainly responsible for the onset of the FE phase when the honeycomb lattice buckles, allowing for an
electrically controllable Rashba-like spin-texture around the K valleys, whose chirality is locked to the polarization direction and therefore fully reversible upon FE switching.
Such Rashba split bands
can be effectively detected by spin-resolved spectroscopic techniques, and the
process of hole and electron injection allows for the engineering of two
dimensional spin field-effect-transistors (FETs) \cite{GeTe}. Even though the spin-valley and Rashba phenomenologies appear to be substantially decoupled, our work suggests a route towards the integration of
valleytronic and spintronic features in FE multivalley materials,
opening unforeseen possibility in the exciting world of spintronics. Currently
the growth of these 2D monolayers on a suitable substrate is difficult, 
and the effect of the substrate has been only recently addressed \cite{zhuang-sub}. 
Nevertheless, it is highly desirable to find ways of preparing these fascinating 2D monolayers. 

\section{ACKNOWLEDGMENTS}
Figures are plotted using VESTA and Gnuplot packages. The authors
are thankful to the HPC Center of NCSU for computing resources. A.S. acknowledges support from CNR Short Term Mobility program prot. AMMCNT-CNR0026336 for the visiting stay at  IFW Leibniz Institute, Dresden, Germany where 
this project was conceived and started.


\end{document}